\documentclass[letterpaper,twocolumn,10pt]{article}
\usepackage{usenix2019_v3}

\usepackage{tikz}
\usepackage{amsmath}
\usepackage{pifont}

\usepackage{ifthen}
\usepackage{xspace}
\usepackage{paralist, tabularx}
\usepackage{booktabs} 
\usepackage{caption,subcaption}    
\usepackage{graphics}

\newboolean{publicversion}
\setboolean{publicversion}{false}

\ifthenelse{\boolean{publicversion}}{
  \newcommand{\grumbler}[2]{}
}{
  \newcommand{\grumbler}[2]{\textcolor{red}{\bf #1: #2}}
}

\newcommand{\sys}[0]{ReLayTracer\xspace}

\begin{document}

\date{}

\title{\Large \bf Slicing the IO execution with \sys}

\author{
    {\rm Ganguk Lee}\\
    KAIST\\
    lgw606@kaist.ac.kr
    \and
    {\rm Yeaseul Park}\\
    KAIST\\
    yeaseulpark@kaist.ac.kr
    \and
    {\rm Jeongseob Ahn}\\
    Ajou University\\
    jsahn@ajou.ac.kr
    \and
    {\rm Youngjin Kwon}\\
    KAIST\\
    yjkwon@kaist.ac.kr
} 
%
%


\maketitle


\begin{abstract}


Analyzing IO performance anomalies is a crucial task in various computing environments,
ranging from large-scale cloud applications to desktop applications. However, the
IO stack of modern operating systems is complicated, making it hard to understand
the performance anomalies with existing tools. Kernel IO executions are frequently
interrupted by internal kernel activities, requiring a sophisticated IO profile
tool to deal with the noises. Furthermore, complicated interactions of concurrent
IO requests cause different sources of tail latencies in kernel IO stack.
As a consequence, developers want to know fine-grained latency profile across
IO layers, which may differ in each IO requests.
To meet the requirements, this paper suggests \sys, a per-request, per-layer IO profiler. \sys enables
detailed analysis to identify root causes of IO performance anomalies by providing
per-layer latency distributions of each IO request, hardware performance behavior,
and time spent by kernel activities such as an interrupt.

\end{abstract}


\section{Introduction}

Understanding IO performance problems is challenging. Performance of kernel IO stacks
are affected by underlying hardware behaviors such as CPU cache locality~\cite{densefs,flexsc,Yang:2012}.
The hardware behaviors add an unexpected delay to kernel IO executions, causing
high performance variations and tail latency. Also, kernel IO executions are often
interrupted by internal kernel activities such as interrupts and exceptions or
scheduler preemptions, make it hard for developers to pinpoint the root cause
of an unexpected performance anomaly.

What makes the cases harder is that developers require fine-grained profiling information
from the complex kernel IO stack. Modern IO stack is built by a set of abstraction layers
each of which has different performance characteristics.
Developers want to profile the latency breakdown of each layer to identify performance bottlenecks.
Furthermore, they would like to know latency distributions of each IO request
to understand what request causes a tail latency and which layer causes the slowdown
comparing to other requests.

In response to the requirement of kernel IO profiling, there are many research and
practical tools to make an effort to provide useful information.
They provide latency breakdown of entire kernel IO stack or block layer but
do not give detailed latency distributions of individual requests~\cite{Joukov:2006, Traeger:2008,
Shin:2014, Xu:2015} and require massive kernel changes to collect trace data~\cite{Joukov:2006, Doray:2017, Ruan:2004}.

\begin{figure}
	\center
	\includegraphics[width=0.45\textwidth]
    {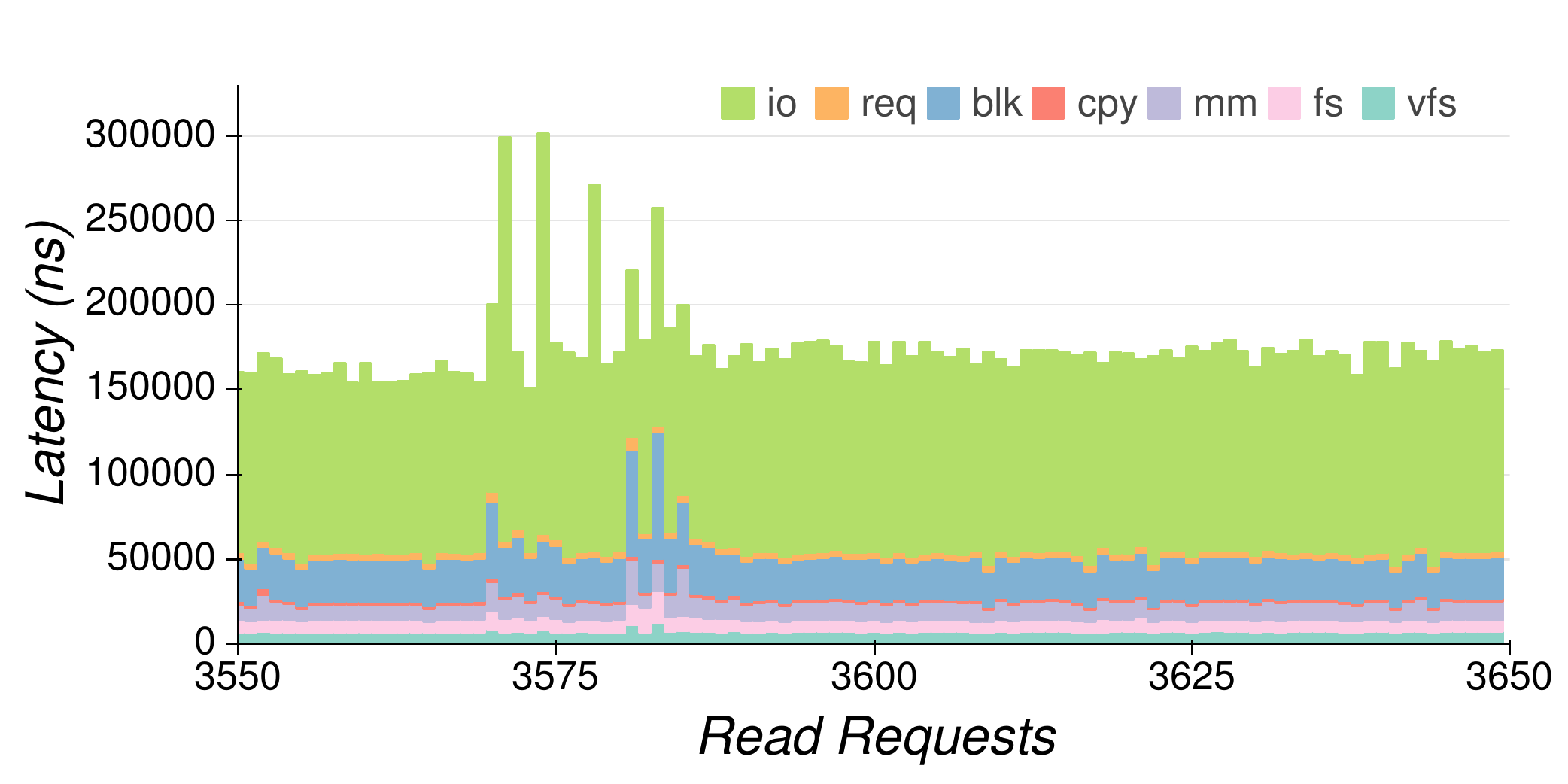}
	\caption{Read latency breakdown of each layer. X-axis is read requests.}
	\label{fig:lat_breakdown}
\end{figure}

In this paper, we introduce a profiling tool for analyzing kernel IO performance in detail,
called \sys ({\tt per-\underline{Re}quest per-\underline{Lay}er} tracer).
\sys provides the latency distributions of each kernel request along with
fine-grained information such as {\it per-abstraction-layer} latency breakdown.
To that end, \sys maps each kernel request (e.g., system call)
to {\it request ID} and tracks the {\it request ID} across the abstraction layers.
By tagging each IO request with a {\it request ID}, which propagates across IO layers, \sys can
report a latency breakdown of each layer of an individual IO request.
Figure~\ref{fig:lat_breakdown} shows the latency breakdown of IO layers profiled by \sys.
Peaks show tail latency, and Figure~\ref{fig:lat_breakdown} shows where the latency peaks happen among
the seven layers. In Linux, a background kernel thread performs device IO asynchronously.
\sys traces the {\it off-CPU event} made by an IO request using the {\it request ID}.
To provide a precise latency breakdown, \sys analyzes an unexpected
delay made by internal kernel activities and accounts them separately.
Also, \sys monitors hardware performance behavior (e.g., IPC) along with a latency profile of each layer,
supporting reasoning about tail latency.

\sys adopts the split architecture consisting of a front end and a back end. The front end of \sys
runs with the target system to profile, collecting data with minimal runtime overhead.
The back end of \sys, working as separate processes, processes data collected by the front end and visualizes the
processed data with graphs. The front end of \sys leverages the dynamic instrumentation
framework~\cite{lttng,ebpf,dtrace,stap,ftrace}, supported in the most of modern operating systems, to
minimize profiling overheads ({\it low overhead}) by tracing only required execution points of interest.
With the dynamic instrumentation, \sys can
profile any kernel subsystems ({\it versatility}) and easily adapt kernel code changes ({\it portability}).
\sys provides fine-grained performance profiling with 3 - 10\% for random read, 0.1\% for sequential read runtime overhead.

This paper makes the following contributions:
\begin{compactitem}
\item By tracing each IO request with a {\it request ID}, \sys can trace per-layer latency on the path
of executing an individual IO request.
\item By monitoring internal kernel activities, \sys can separately account the interference caused the
kernel activities.
\item By bundling software and hardware performance information, \sys can precisely analyze the Linux
IO performance.
\end{compactitem}
The current scope of the work focuses on the read path of IO. Adding support for
other IO system calls is future work.



\section{Design and Implementation}

\begin{figure}
\center{\includegraphics[width=0.4\textwidth]
	{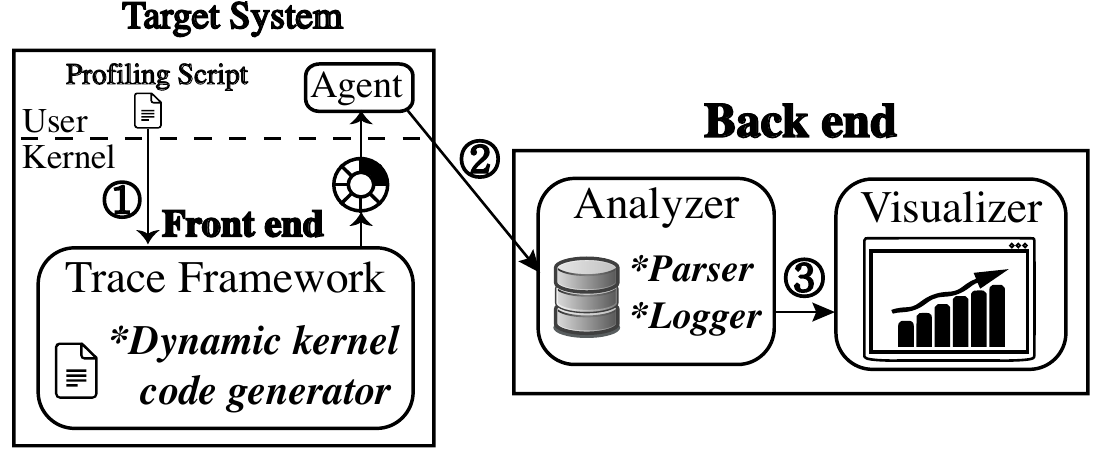}}
	\caption{\label{fig:outline} \textbf{\sys} Architecture.}
\end{figure}

The goal of \sys is to provide precise and fine-grained latency information of kernel IO subsystems
with minimal performance overhead.
\sys starts tracing on each system call and generates trace data profiling timing of
kernel IO layers. With the trace data, \sys computes fine-grained latency breakdown of each IO layer.
To minimize runtime overhead of a target system, \sys leverages an existing lightweight
instrumentation framework~\cite{kprobe,uprobe, tracepoints, ftrace,perf, ebpf, lttng,stap,dtrace}. 
The number of trace points directly affects the runtime
overhead of a target system. Kernel instrumentation frameworks have the flexibility
to control the number of trace points, enabling \sys to manage the runtime overhead.
The remainder of this section describes the design of \sys by giving a case study
of the read system call.

\subsection{Architecture}
\sys comprises a front end and a back end system. Figure~\ref{fig:outline} shows
the overall architecture of \sys. The front end is planted to a target system
and collects tracing data, and the back end processes data obtained from the front end and visualizes them.

\begin{figure}
\center{\includegraphics[width=0.4\textwidth]
	{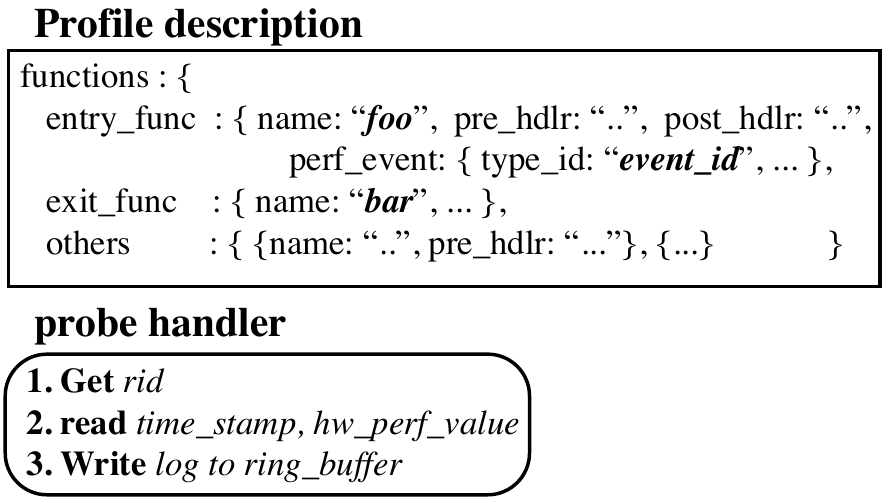}}
	\caption{\label{fig:frontend} Profile script describes probing points and probe handers. A probe
	handler generally consists of getting rid, collecting trace data to a log, and writing the log
	to the ring buffer.}
\end{figure}

\textbf{Leveraging instrumentation.} Modern operating systems support kernel dynamic instrumentation~\cite{lttng,ebpf,stap,dtrace,ftrace}.
A kernel dynamic instrumentation allows execution of user-defined code into in-memory kernel code
at function boundaries (i.e., entering and exiting a function) or in any lines (called {\it probe point}).
When a kernel execution hits an installed probe point, the instrumentation framework calls
a procedure (called {\it probe handler}) registered by the framework.
The front end takes advantage of existing kernel dynamic instrumentation frameworks to
install arbitrary probe points into a running kernel~\cite{kprobe, uprobe, tracepoints, ftrace, perf,stap}.
The front end takes a profile script to install probe points.
In Figure~\ref{fig:frontend}, we show an example of the profile script.
The profile script is written in a domain-specific language
describing probe points of each kernel layer as a list of kernel functions, entry and exit
actions defined as a probe handler, and hardware events~\cite{perf} to monitor.
The tracing framework compiles the profile script to a binary and inserts it
into kernel memory using a kernel dynamic instrumentation framework.

\subsection{Front end}

The primary goal of the front end install probe points and probe handlers
to collect timing information of each IO layer in each IO request.
The front end installs probe points in execution paths of the system call,
off-CPU events, and interrupt service routines.

\begin{figure}
\center{\includegraphics[width=0.45\textwidth]
	{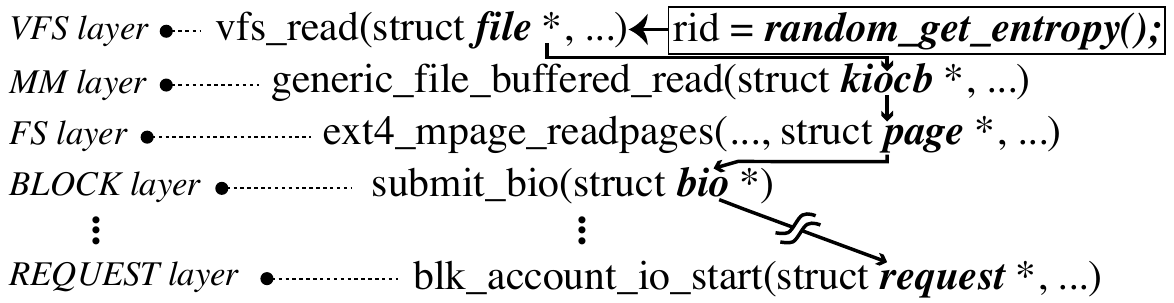}}
	\caption{\label{fig:RidCallChain} request id propagation through function calls.
	After the request is queued at the request layer, SCSI driver keeps checking the queue and handles the queued requests. }
\end{figure}

\noindent \textbf{Tracing system calls.}
Because the system call is the way of entering into the kernel space,
we first focus on how we trace system calls to get the {\it per-request} latency distribution.
An IO system call extends across multiple layers of kernel components.
For example, in Linux, a read system call travels through
VFS and page cache layer, file system layer, block IO layer, and device driver layer.
Each layer has different performance characteristics such as memory copy bandwidth or
slowdown by lock contentions. In addition to that, concurrent system calls make the analysis more complicated because
it is hard to distinguish what system call is executing without having
a complex function parameter analysis and matching thread id which causes
non-negligible overhead in IO-intensive workloads.

To provide lightweight and fine-grained tracing, \sys associates a {\it request id}
(or {\it rid}) with an IO system call to trace. {\tt rid} is a timestamp monotonically
increasing and serves as a unique identifier of an IO system call. To track
individual IO system calls across the layers, \sys propagates an assigned {\tt rid} for each read system call
to lower layers of the kernel as depicted in Figure~\ref{fig:RidCallChain}. The value of {\it rid} is initially stored
in {\tt struct file} (VFS layer), transferred to {\tt struct kiocb} (memory management layer)
and the subsequent layers. To implement the {\it rid} propagation, we slightly modify kernel data structures
to hold the {\it rid} and add code to maintain the {\it rid} value when a layer is switched.
We use a kernel instrumentation framework to change 29 lines of code in the Linux kernel for tracing read IO 
(7 lines for {\it rid} in struct and 22 lines for {\it rid} propagation). 
This minimizes performance and space overhead when tracing a large volume of IO requests.
\sys can work with any instrumentaion framework that supports on-the-fly
modification of kernel code.

\noindent \textbf{Tracing off-CPU events.} In addition to system calls, \sys supports profiling off-CPU events.
Some IO system calls execute their part of code paths on a different CPU,
which makes it hard to distinguish the origin of an IO request.
For example, read or write system calls make a block IO request
to a kernel IO thread and wait until the kernel IO thread completes the IO request.
The off-CPU execution path must be added to the total latency of the system calls.
To associate a system call and off-CPU execution, \sys transfers the {\it rid} to
off-CPU kernel functions.
Generally, there is a shared data structure for the off-CPU handler (e.g., {\tt struct bio}
and {\tt struct request} in Figure~\ref{fig:RidCallChain}).
\sys installs probe points on off-CPU kernel events to profile and
delivers {\it rid} to an off-CPU probe point via a data structure used for a parameter.

\noindent \textbf{Tracing kernel activities.} Tracing only execution paths of system calls
causes an inaccurate profile result because the kernel execution can be interrupted by
various external events. For example, an interrupt handling during a read
system call causes a delay in a layer. If probe points are installed at function boundaries
of a system call execution path, the delay would not be accounted.
To sort out the delay made by the kernel activities, \sys also installs probe points
on interrupt handling routines, schedule in and out handlers, and the wakeup points
of background kernel threads, recording occurences of the event and its timing
information.


\sys installs a probe handler, described in Figure~\ref{fig:frontend}, on each probe point.
A probe point can be either an entry and an exit of a function or both based on
profile description in Figure~\ref{fig:frontend}.
A probe handler creates a trace log consisting of {\tt <function name, process id, cpu id,
rid, timestamp, hardware events>} and records the log to a ring buffer. The ring buffer
is a protected, shared memory between kernel and user-level applications.
The ring buffer size is 4MB (1024 * 4KB page) by default. A \sys's
user-level agent in a target system periodically sends trace logs to the back end
via the network to avoid buffer overruns.

\subsection{Back end}
The goal of the back end is to compute {\it per-layer} latency breakdowns from
raw log records traced by the front end.
The back end aggregates the log records with a {\it request id} and then
invokes an analyzer for each {\it request id}. 
Along with a profile description, the analyzer takes a layer description, which
specifies layer types and how to separate layers with given function names.
With the layer description, the analyzer computes {\it per-layer} latency of each
system call request which has the same {\it rid} in the log records. If a traced function, $f_1$,
calls a function, $f_2$, belonging to a lower layer, the analyzer subtracts the
time taken by $f_2$ when computing latency of $f_1$. The analyzer checks
whether an off-CPU event happens in the middle of execution of a traced function.
If an event happens, the analyzer separately accounts the latency caused by
the off-CPU event when computing latency of a traced function. In case that
a hardware event (e.g., CPI) is recorded in a traced function, the analyzer correlates the hardware event with
the traced function and records the hardware event data with a computed latency.
Finally, the analyzer stores the processed data to an intermediate format, and
the visualizer draws a figure, which can be rendered in web browsers.


\section{Evaluation}

In this section, we explore possible application scenarios of \sys to understand kernel behaviors at fine-grained level.
Our prototype leverages the {\tt ebpf} framework supported by standard Linux to make use of the tracing facility~\cite{ebpf}.
The target system consists of the Intel Xeon CPU E5-2695 v4 CPU, 128GB DDR4 memory, and 256GB SATA SSD.
We use a Ubuntu 16.04 machine with modified kernel version 4.15 and the ext4 filesystem.
We run various workloads of synchronous read I/O operations generated by FIO benchmark \cite{fio} on the target system
to observe the effectiveness of \sys in terms of the {\it per-request} and {\it per-layer} profiling facility.
First, we examine single-threaded sequential and random read access patterns, then we move on to the multi-threaded benchmark study.

\subsection{Read System Calls and Page Cache}
In general, when the Linux operating system receives a read request from a user, it first looks up
in the main memory cache, called \textit{page cache}, to check if the requested data is already in memory.
Based on this mechanism, we categorize the read system requests into
two classes - page cache hit and page cache miss.
In cases of page cache hits, the read requests are rapidly served without
traversing through multiple kernel layers.
On the other hand, read system calls that experience page cache misses are more time-consuming because
they go down to lower kernel layers to access the storage medium and
bring the requested page into the memory.

\begin{figure}[]
\centering
\includegraphics[width=0.5\textwidth]{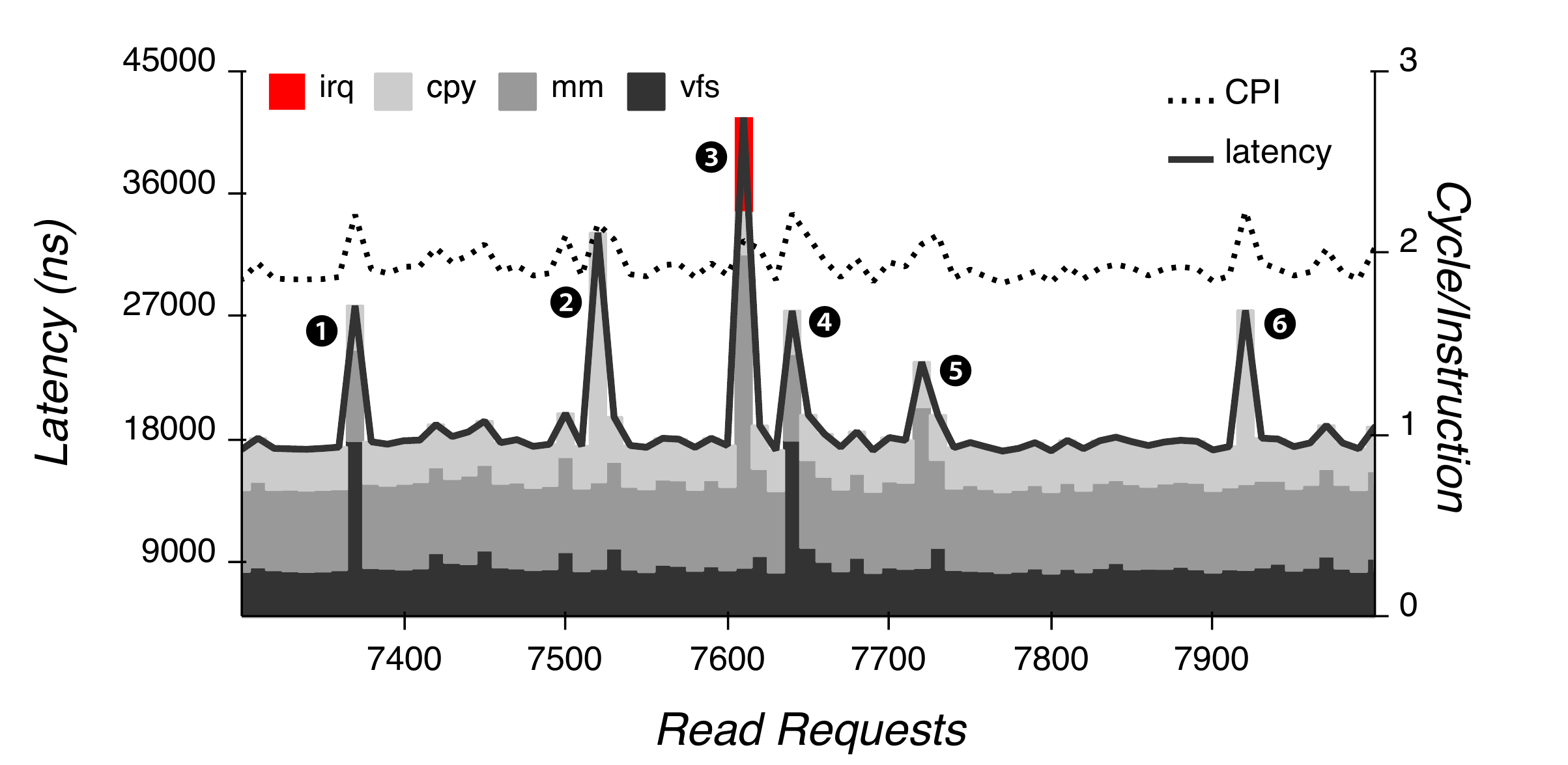}
\caption{{\it Per-request} latency (solid) and CPI (dotted) for single-threaded sequential read requests in chronological order.
  Only page cache hit cases are shown. Each latency is broken down into four kernel layers - {\tt vfs, mm, cpy, irq}}
\label{fig:latencyandCPI}
\end{figure}

\noindent {\bf Tail latency within page cache hits.}
We expect that the latency variation among read requests would not be significant if the requested data reside in the page cache.
Interestingly, however, \sys reports that latency still varies widely within page cache hit cases.
Figure~\ref{fig:latencyandCPI} shows {\it per-request} read system call latency (for page cache hits) when a file is read sequentially by a single thread.
The x-axis represents individual requests in chronological order.
For most requests, the latency is measured around 18000ns, but we observe that there are several noticeable spikes that rise above 25000ns.
To identify the cause of tail latencies, we additionally analyze the CPI
(cycles per instruction) provided by \sys front end for each request, which is depicted by the dotted line in the same figure.
As the latency fluctuates, the CPI follows the same trend.
Using the {\it per-layer} facility of \sys, we dissect the latency into four layers
 - virtual file system ({\tt vfs}), memory management ({\tt mm}), page copy ({\tt cpy}), and interrupt request handle ({\tt irq}).
The data visualization output by our tracer shows that the fluctuation does not occur in a particular layer,
but, instead, different components are responsible for slowdown in each request.
{\tt vfs} layer turns out to be the causative factor of tail \ding{202} and \ding{205}.
Similarly, {\tt cpy} layer was the bottleneck in tail \ding{203} and \ding{207}.
The layer breakdown utility shows us that tail \ding{204} additionally contains {\tt irq} layer.
In fact, the high latency in this particular request was due to a hardware interrupt for IO completion which occurred at {\tt mm} layer.
{\tt irq} not only adds an extra layer in the request
but also introduces huge context switch overhead in the interrupted layer.
\sys allows in-depth analysis to locate unusual latency spikes even within page cache hits
and further diagnose the root cause of each at hardware and software levels.

\begin{figure}[]
\centering
\includegraphics[width=0.4\textwidth]{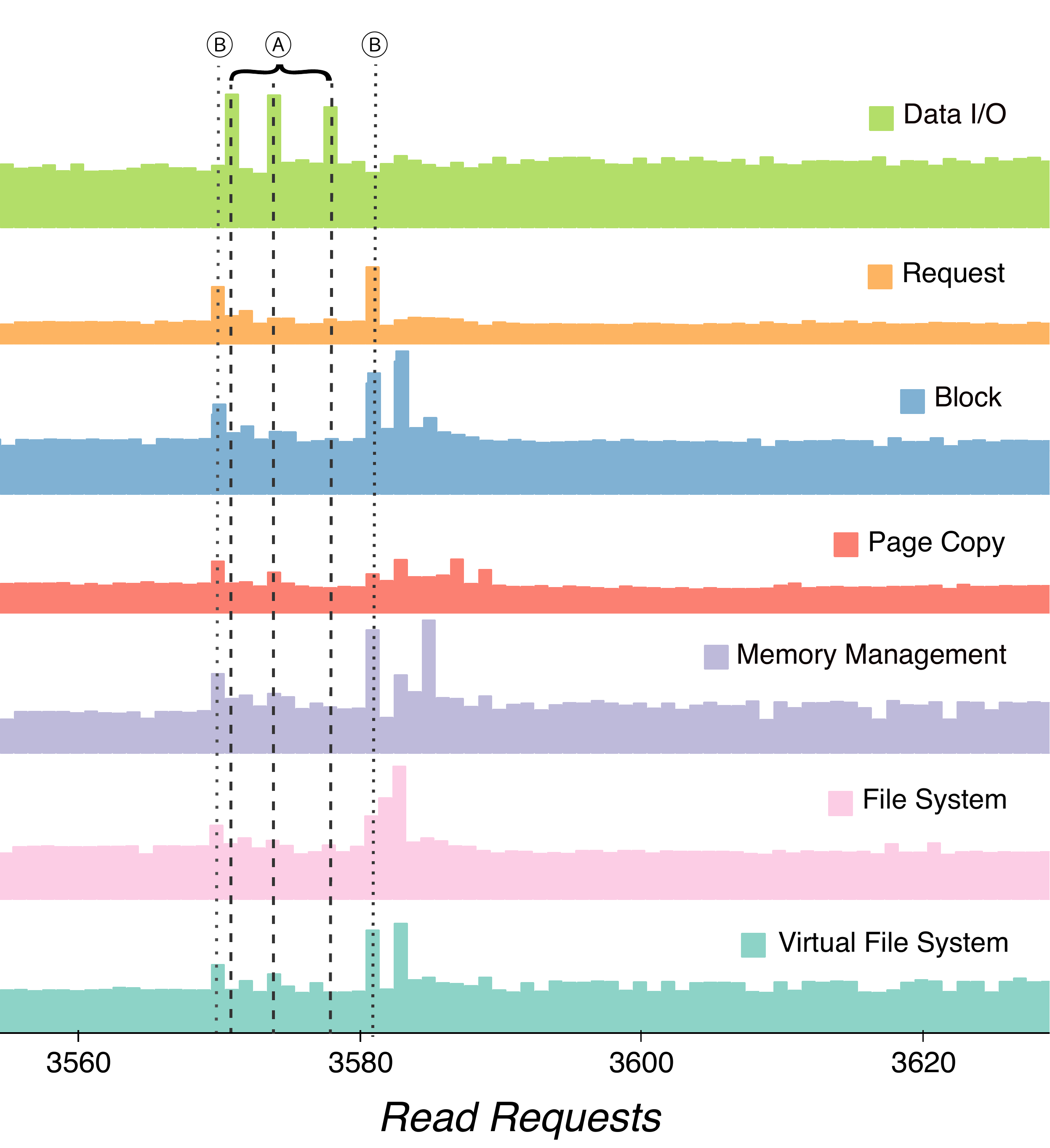}
\caption{{\it Per-layer} breakdown of latency for single-threaded random read requests in chronological order.}
\label{fig:per_layer}
\end{figure}

\noindent {\bf Tail latency within page cache misses.}
We also examine a scenario where the read system calls do not benefit from the page cache due to the random access pattern.
Figure~\ref{fig:per_layer} shows {\it per-layer} breakdown of request latency for single-threaded random reads on a file.
When read requests are made for random positions in a file, most requests experience page cache misses
and suffer much higher latency because they penetrate through deeper layers down to storage to bring data into memory.
A large portion of the time is spent in performing disk I/O, as depicted by the
thick green layer ({\tt io}) in Figure~\ref{fig:lat_breakdown}.
Figure~\ref{fig:per_layer} corresponds to the layer decomposition of requests in the spiking region (requests 3560 - 3600) of Figure~\ref{fig:lat_breakdown}.
By breaking each request down to units of a kernel layer,
it is possible to observe that each burst is caused by different layers
even within a small interval of 40 requests.
The largest three spikes (\textcircled{\raisebox{-0.8pt}{A}}) are due to delay in the Disk I/O layer,
but the smaller spikes that appear immediately before and after (\textcircled{\raisebox{-0.8pt}{B}}) are caused by
equally proportional slowdown in all other six layers.
Since the random read requests do not have access locality, we anticipate that there is no significant latency difference among them.
However, \sys informs us that multiple tail latencies can occur in a batched manner within a small time interval
and that different layers contribute to each tail.
\subsection{Profiling Multi-threaded Behaviors}

\begin{figure}[]
\centering
\includegraphics[width=0.5\textwidth]{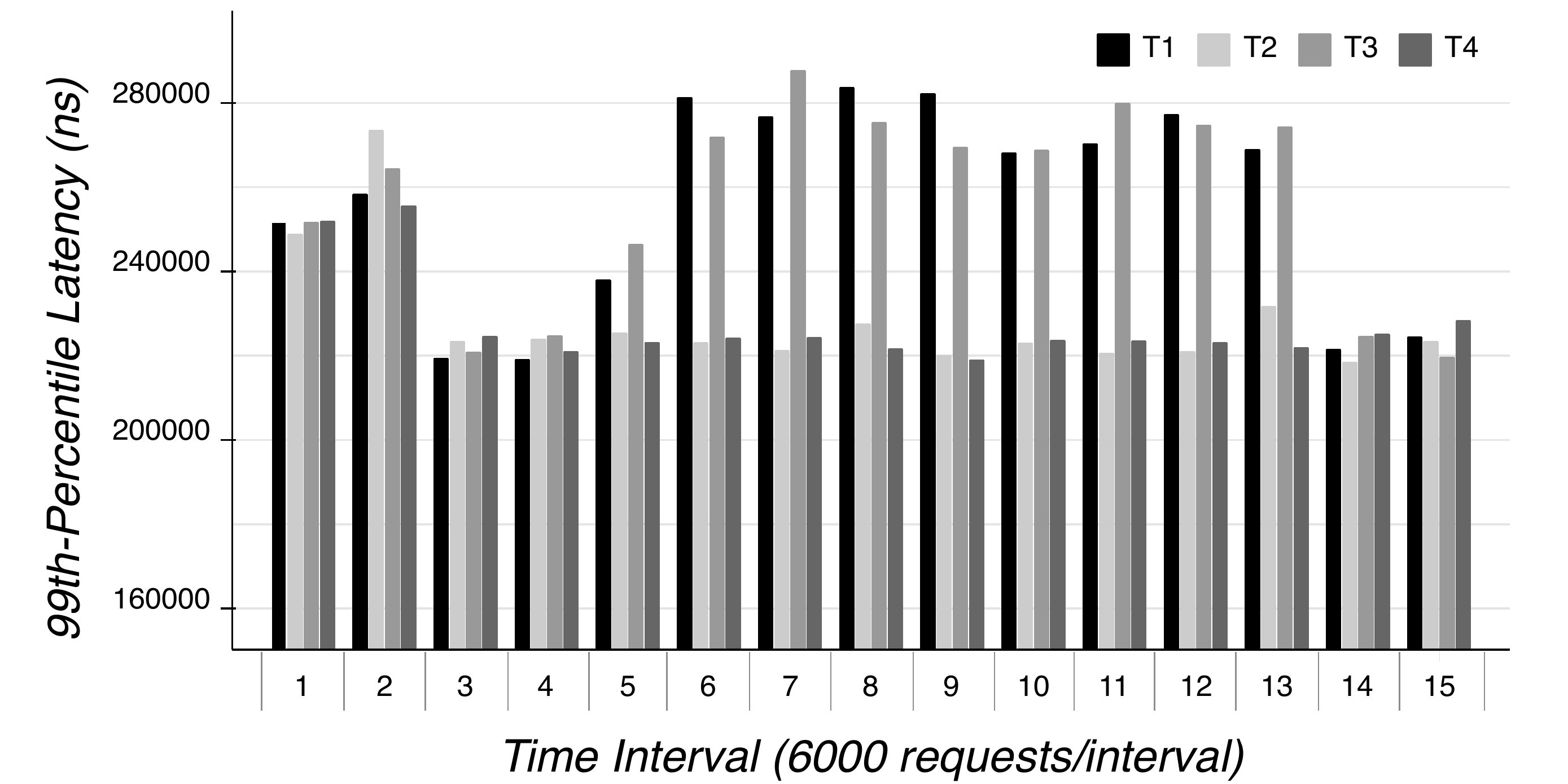}
\caption{99th-percentile latency within each time interval for threads T1, T2, T3, T4
  simultaneously performing random read operations on a single file.
  Total 90,000 requests are divided into 15 intervals, each containing 6000
requests.}
\label{fig:FairnessBreakdown}
\end{figure}

In this subsection, we examine the {\it fairness} of issuing random read operations by multiple threads.
As each thread makes its read request, the requests end up mixed across the underlying kernel layers.
With its ability to measure the latency on a {\it per-request} basis, however, \sys is able to retrieve the owner thread of each request.
Figure~\ref{fig:FairnessBreakdown} shows 99th percentile latency of each thread over time with 6,000 requests as a group of the time interval ({\it window}), with 
each bar representing individual threads. 
From window 5 to 13, we observe that T1 and T3 experience much longer tail latency.
In fact, we identify that the burden of IRQ handling was unfairly distributed
throughout threads. Certain threads served much more IRQ requests than the others and had to spend up to 
60x more time on IRQ handling as shown in Table~\ref{tab:irq_time}.


\begin{table}[]
\resizebox{\columnwidth}{!}{%
\begin{tabular}{c|r|r}
\toprule
Thread  & \# of IRQ handled &  Time spent on IRQ (ns) \\
\midrule
T1        & 7,276              & 47,849,914        \\
T2        & 128                & 816,828           \\
T3        & 2,953              & 19,398,496        \\
T4        & 121                & 787,582           \\
\bottomrule
\end{tabular}
}
\caption{Number of IRQ events handled by each thread and the total time spent on handling IRQ during execution.}
\label{tab:irq_time}
\end{table}

\subsection{Overhead}
We end the section by evaluating the performance overhead of \sys.
First, we measure the read throughput (bandwidth) for sequential read access patterns with 8 threads.
In this case, threads do not experience any throughput degradation when running with \sys.
Since most of the read operations are served by the page cache, the depth of IO path for most requests is small.
As a result, the overhead introduced by the additional CPU cycles from the
tracer is negligible. 

Next, we measure the overhead of \sys for random read accesses by 8 threads.
Table~\ref{tab:overhead} shows the throughput overhead with increasing depth of probe layers.
As the probe depth increases, the number of probe points also increases.
For random accesses, the page cache does not help, so all possible tracing points across the 8 layers will be reached frequently,
resulting in non-negligible overhead.
At minimum depth ({\tt L1}), the bandwidth degradation is only 3\%, but it can increase up to 10.7\% with maximum depth ({\tt L8}).
This result depicts the tradeoff between the profiling depth and the
overhead.
In response to this, \sys provides a control knob to users to adjust the profiling granularity.

\begin{table}[]
\resizebox{\columnwidth}{!}{%
\begin{tabular}{l|cccccccc}
\toprule
Probe depth     & L1 & L2  & L3 & L4  & L5  & L6  & L7  & L8   \\ \hline
Overhead(\%) & 3  & 3.3 & 6  & 6.1 & 8.3 & 8.4 & 9.5 & 10.7 \\ \hline
\# Probe points  & 2  & 6   & 8  & 9   & 12   & 13   & 14   & 15    \\
\bottomrule
\end{tabular}%
}
\caption{Bandwidth degradation with increasing probe depth when tracing
8 threads simultaneously performing random read a total of 2GB of data.}
\label{tab:overhead}
\end{table}


\section{Conclusion}
To identify the root causes of performance anomalies of kernel IO, \sys provides
fine-grained performance profiling of kernel IO stacks. \sys tags each IO request
with {\tt request ID} and traces the {\tt request ID} through IO layers, measuring
precise latency of each IO layer the IO request travels. \sys leverages existing
kernel instrumentation framework to dynamically install trace points and collects
latency information within 3 - 10\% overhead.

%


\bibliographystyle{plain}
\bibliography{biblio}

\end{document}